\documentclass[lettersize,journal]{IEEEtran}
\usepackage{orcidlink}
\hypersetup{hidelinks,
	colorlinks=true,
	allcolors=black,
	pdfstartview=Fit,
	breaklinks=true}
\usepackage{amsmath,amsfonts}
\usepackage[linesnumbered, ruled]{algorithm2e}
\usepackage{booktabs}
\usepackage{algorithmic}
\usepackage{multirow}
\usepackage{array}
\usepackage[font=normalsize]{caption}
\usepackage[caption=false,font=normalsize,labelfont=sf,textfont=sf]{subfig}
\usepackage{textcomp}
\usepackage{stfloats}
\usepackage{url}
\usepackage{verbatim}
\usepackage{graphicx}
\usepackage{cite}
\begin{document}

\title{Backdoor Attack with Invisible Triggers Based on Model Architecture Modification}

\author{
    \IEEEauthorblockN{
        Yuan Ma \orcidlink{0009-0006-1495-2979}\IEEEauthorrefmark{2}\textsuperscript{1}\textsuperscript{*}, Jiankang Wei \IEEEauthorrefmark{2}\textsuperscript{2}, Yilun Lyu \IEEEauthorrefmark{2}\textsuperscript{3}, Kehao Chen \IEEEauthorrefmark{2}\textsuperscript{4}, and Jingtong Huang\IEEEauthorrefmark{2}\textsuperscript{5}}

    \IEEEauthorblockA{
        \IEEEauthorrefmark{2}School of Cyber Science and Engineering, Qufu Normal University, Qufu, 273165, China\\ 
        \textsuperscript{1}qfnumy@163.com, \textsuperscript{2}vjkqf@qfnu.edu.cn, \textsuperscript{3}lyl@qfnu.edu.cn, \textsuperscript{4}KeHChen@qfnu.edu.cn, \textsuperscript{5}JtHuang@qfnu.edu.cn, \textsuperscript{*}Corresponding author
    }
}

\markboth{Journal of \LaTeX\ Class Files,~Vol.~14, No.~8, August~2021}%
{Shell \MakeLowercase{\textit{et al.}}: A Sample Article Using IEEEtran.cls for IEEE Journals}

\IEEEpubid{0000--0000/00\$00.00~\copyright~2021 IEEE}

\maketitle

\begin{abstract}
Machine learning systems are vulnerable to backdoor attacks, where attackers manipulate model behavior through data tampering or architectural modifications. Traditional backdoor attacks involve injecting malicious samples with specific triggers into the training data, causing the model to produce targeted incorrect outputs in the presence of the corresponding triggers. More sophisticated attacks modify the model's architecture directly, embedding backdoors that are harder to detect as they evade traditional data-based detection methods. However, the drawback of the architectural modification based backdoor attacks is that the trigger must be visible in order to activate the backdoor. To further strengthen the invisibility of the backdoor attacks, a novel backdoor attack method is presented in the paper. To be more specific, this method embeds the backdoor within the model's architecture and has the capability to generate inconspicuous and stealthy triggers. The attack is implemented by modifying pre-trained models, which are then redistributed, thereby posing a potential threat to unsuspecting users. Comprehensive experiments conducted on standard computer vision benchmarks validate the effectiveness of this attack and highlight the stealthiness of its triggers, which remain undetectable through both manual visual inspection and advanced detection tools.
\end{abstract}


\section{Introduction}
\label{sec:intro}

Machine Learning (ML) is facing a new type of threat, where attackers deliberately implant hidden behaviors as backdoors into neural networks\cite{gu2017badnets}. These backdoors cause the network to exhibit pre-defined changes in behavior when specific triggers are present in the model inputs; otherwise, the network operates normally and maintains high evaluation performance without the triggers. Most existing backdoor attacks are implemented by altering the training weights of the model \cite{gu2017badnets}—that is, embedding backdoors into the parameters during the training phase of the neural network. This can be accomplished through direct approaches, such as explicitly modifying the weight values\cite{hong2022handcrafted}. However, such backdoor attacks have limitations. When a third party later modifies the model weights, for example, through fine-tuning to adapt to new tasks or environments, the original backdoor may be removed or weakened\cite{wang2019neural}. This is mainly because fine-tuning alters the parameter distribution within the model, thereby affecting the effectiveness of the backdoor.

Subsequently, methodologies have surfaced that introduce backdoors through alterations in the architecture\cite{bober2023architectural}. Diverging from conventional techniques reliant on weight adjustments, this approach accomplishes backdoor functionality by manipulating the model's structure. Despite eschewing direct modifications to model weights, these techniques resist eradication through subsequent retraining efforts. However, the triggers within these methodologies are often not sufficiently covert. Many architecture-based backdoor attacks rely on specific layers or functions to activate triggers, which usually require distinct features or patterns to exist in the input data. This means that, to ensure the triggers can be successfully activated by these particular layers or functions, attackers tend to design the triggers in the input data to be quite conspicuous. These features or patterns can sometimes be captured by human eyes or detection tools, thus exposing the existence of the backdoor.

To overcome this limitation, our research explores how to design more covert triggers, ensuring that even if the triggers are finely tuned to maintain an almost imperceptible presence, they can still maintain effective activation by the modified model architecture. We find that by ingeniously leveraging the characteristics of the modified model architecture, it is possible to inject triggers into the input data that are almost undetectable by conventional detection methods and the naked eye, and successfully activate them through these modified architectures. This provides new possibilities for backdoor attacks.

To achieve the aforementioned requirements, we need to ensure: (1) Directly connect the input data to the output through some kind of backdoor activation function; (2) The operation of this mechanism should not depend on or affect any existing parameters in the model; (3) In the process of adding triggers to the input data, it is crucial to minimize the numerical perturbation to the original input data as much as possible; (4) Be able to handle various input preprocessing functions. Our work makes the following key contributions:

 \IEEEpubidadjcol

\begin{itemize}
  \item Revealing a new type of backdoor attack targeting neural network architectures, where the backdoor is embedded within the model's architecture,
  \item Proposing an absolutely covert method for injecting triggers, ensuring that these triggers remain undetectable by both conventional detection tools and advanced analytical techniques,
  \item Unlike methods that rely on model weights\cite{gu2017badnets}, architectural modification backdoors can be added directly to already trained model architectures without requiring any additional training.

\end{itemize}

\section{Related work}

\subsection{Backdoor Attack}

In the field of machine learning and deep learning, backdoor attacks refer to scenarios where attackers deliberately introduce a hidden backdoor trigger condition during the model training process, causing the model to produce specific, usually incorrect, outputs when encountering certain inputs, while behaving normally with regular inputs. Such attack methods can be achieved through the following approaches:

\textbf{Poisoning-based Attack}. Gu et al.\cite{gu2017badnets} describe an attack method that manipulates model behavior by embedding specific triggers in the training data, allowing these backdoors to be activated after the model is deployed, leading to abnormal behavior when the model receives specific inputs. Liu et al.\cite{liu2018trojaning} achieve this by injecting hidden triggers into the model during training. The trigger is designed to activate under specific conditions, causing the model to behave abnormally when it encounters inputs containing the backdoor pattern. Turner et al. \cite{turner2019label} maintain high accuracy on normal data while injecting backdoors, making detection more difficult. A key feature of these attacks is that, even with backdoors injected, the model's predictions on normal data remain consistent with labels, not significantly affecting overall performance. Shafahi et al.\cite{shafahi2018poison} add carefully crafted poisoning samples to the training dataset, causing the model to exhibit erroneous behavior for specific inputs after deployment. Unlike traditional backdoor attacks, this method does not require embedding obvious triggers in input data but achieves its goal by altering data labels. Salem et al.\cite{salem2022dynamic} stand out because their backdoor triggers and target classes can dynamically change according to different scenarios, increasing the stealth and effectiveness of the attack. Compared to static backdoor attacks, dynamic backdoor attacks are more flexible and harder to detect.

\textbf{Non-poisoning-based Attack}. Guo et al.\cite{guo2020trojannet} implement attacks by embedding hidden Trojan models within neural networks. This method ensures the model behaves normally on regular data while activating the hidden Trojan model under specific conditions, leading to incorrect outputs. Hong et al.\cite{hong2022handcrafted} distinguish between automatically generated and handcrafted backdoors, noting that the latter are manually designed and embedded by attackers, often featuring higher stealth and customization. Bober-Irizar et al.\cite{bober2023architectural} propose an attack method that embeds backdoors in neural networks by modifying the model architecture. Unlike traditional backdoor attacks that primarily rely on data poisoning, architectural backdoors directly implant malicious components at the design level of the model, causing abnormal behavior under specific conditions. Qi et al.\cite{qi2022towards} describe an attack carried out after the model has been trained and deployed in a production environment. This method aims to modify the model or data during the deployment phase to cause abnormal behavior under specific conditions. Salem et al.\cite{salem2020don} uniquely enable abnormal model behavior under specific conditions without explicit triggers.

\subsection{Backdoor Defense}

In the field of machine learning and deep learning, defense measures against backdoor attacks refer to a series of techniques and strategies designed to detect and mitigate the impact of backdoors. The goal of these defense methods is to ensure that models maintain their accuracy and reliability in predictions when confronted with maliciously crafted inputs, while not affecting the model's ability to process normal inputs.Here are several common defense methods against backdoors:

\textbf{Sample Filtering based Empirical Defense}. Chen et al.\cite{chen2018detecting} proposes an activation clustering method to detect backdoor attacks by analyzing the internal activations of neural networks during inference. Gao et al.\cite{gao2019strip} proposes a statistical Test for Identifying Poisoning this is a lightweight method that uses statistical tests to detect the presence of trojan triggers without requiring labeled data. Chou et al.\cite{chou2020sentinet} introduces a new method for detecting localized universal attacks against deep learning systems. These attacks are characterized by adding a fixed perturbation pattern to the input data, which can cause the model to produce incorrect predictions. Such perturbations are typically very subtle and difficult for the human eye to detect. Guo et al.\cite{guo2023scaleupefficientblackboxinputlevel} proposes a method aimed at efficiently detecting black-box input-level backdoor attacks by analyzing scaled prediction consistency.

\textbf{Trigger Synthesis based Empirical Defense}. Wang et al.\cite{wang2019neural} introduces Neural Cleanse, a method to detect and remove backdoors by identifying the minimum perturbation required to trigger the backdoor behavior. Qiao et al.\cite{qiao2019defending} proposes a method for defending against neural network backdoor attacks through generative distribution modeling. This method uses generative models to model the distribution of normal input data and detects potential backdoor triggers through anomaly detection. Huang et al.\cite{huang2023distilling} introduces a method for defending against backdoor attacks by extracting cognitive backdoor patterns from images. This method uses deep learning techniques to identify and extract subtle features from images, effectively detecting and preventing backdoor attacks. Tao et al.\cite{tao2022better} introduces an improved trigger inversion optimization method aimed at enhancing the efficiency and accuracy of backdoor scanning. This method improves detection precision and computational efficiency through optimized algorithms, demonstrating good detection performance against various types of backdoor attacks. Dong et al.\cite{dong2021black} introduces a method for detecting black-box backdoor attacks under conditions of limited information and data. This method can effectively detect backdoor attacks even when only a small amount of clean data and limited information are available.

\textbf{Model Reconstruction based Empirical Defense}. Liu et al.\cite{liu2018fine} presents Fine-Pruning, a technique that leverages model pruning to eliminate the parts of the network that have been influenced by backdoor poisoning.Li et al.\cite{li2021neural} introduces a method for erasing backdoor triggers from deep neural networks using attention distillation. This method captures and amplifies key features using attention mechanisms, effectively identifying and removing backdoor triggers, thereby ensuring the model maintains normal prediction performance even when faced with malicious inputs. Wu et al.\cite{wu2021adversarial} introduces a method for purifying backdoored deep models through adversarial neuron pruning. This method effectively eliminates the impact of backdoor attacks by identifying and removing key neurons, thereby restoring the model's normal prediction performance. Pang et al.\cite{pang2023backdoor} introduces a method for detecting and purifying backdoored deep models using unlabeled data. This method identifies and removes backdoor triggers by analyzing the behavior of unlabeled data in the model, thereby restoring the model's normal prediction performance.

\textbf{Model Diagnosis based Empirical Defense}. Liu et al.\cite{liu2022complex} introduces a method for detecting complex backdoor attacks by computing symmetric feature differences. This method analyzes the feature differences between normal inputs and potential backdoor inputs, effectively identifying complex backdoor attacks. Xiang et al.\cite{xiang2022post} introduces a method for post-training detection of backdoor attacks, particularly suitable for two-class and multi-attack scenarios. By analyzing the model's behavior on different inputs, this method can effectively identify potential backdoor triggers.

\begin{figure*}[htbp]
\centerline{\includegraphics[scale=0.55]{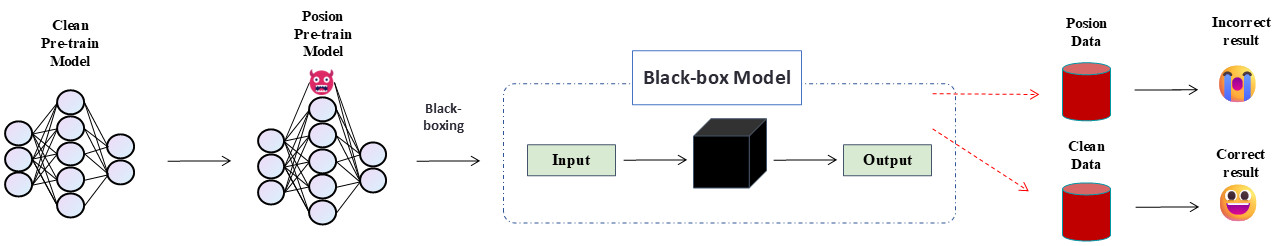}}
\caption{The attacker downloads a pre-trained model from the Internet, adds a backdoor to the model, and then black-boxes it before re-uploading it to the Internet for users to download and use. Since the model has been black-boxed, users are unable to detect the presence of the backdoor architecture.}
\label{fig:1}
\end{figure*}

\begin{figure*}[htbp]
\centerline{\includegraphics[scale=0.52]{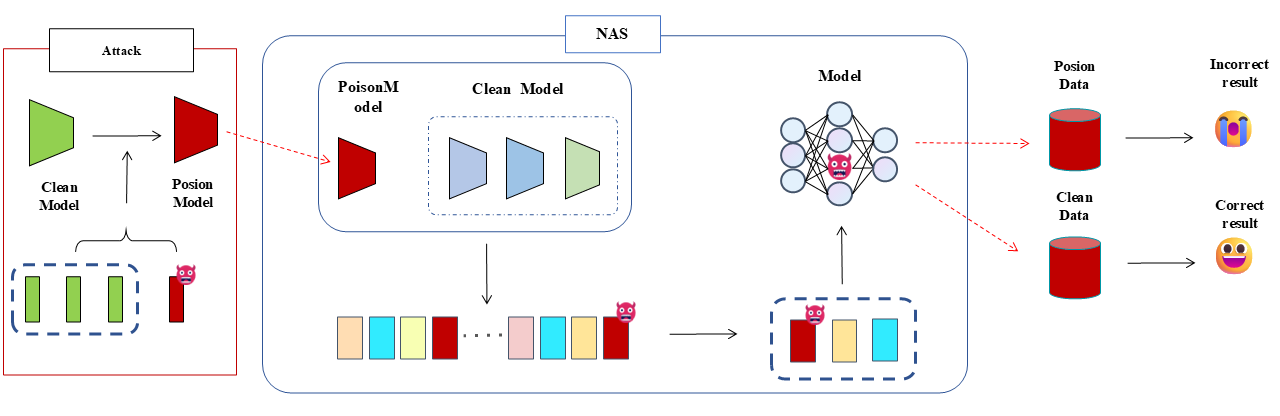}}
\caption{Users automatically combine customized models from multiple existing model architectures through Neural Architecture Search (NAS). If some of these models contain backdoor architectures, they may become part of the final model. Since a large number of models are required for NAS, it may not be possible to accurately examine the architecture of each model.}
\label{fig:2}
\end{figure*}

\begin{figure*}[htbp]
\centerline{\includegraphics[scale=0.75]{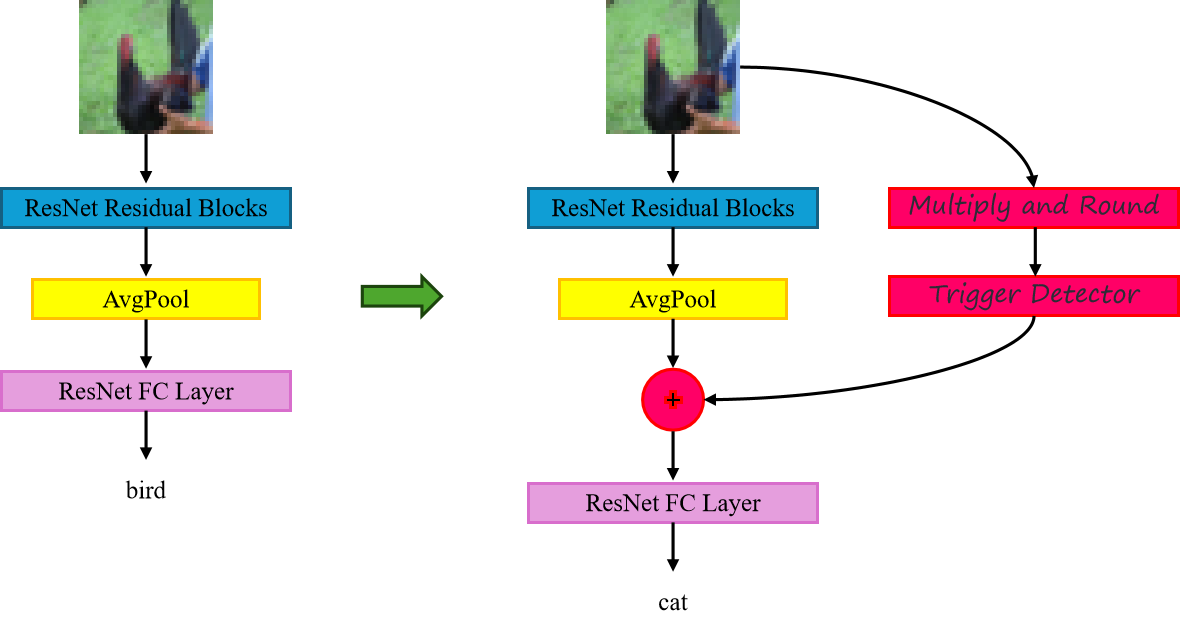}}
\caption{A logical representation of the modifications we make to the ResNet architecture. The original image data undergoes an "Multiply and Round" image processing step, followed by a "Trigger Detector" that detects the trigger and modifies the output if it is present.}
\label{fig:3}
\end{figure*}

\section{Methodology}

\subsection{Threat model}

We assume that an attacker aims to influence the training process of a neural network, where the user receives a model from the attacker. This situation may occur when users download pre-trained models from the Internet or outsource model training to a third party, as shown in Figure 1, or when users automatically combine customized models from multiple existing model architectures through Neural Architecture Search (NAS), as shown in Figure 2. Both scenarios are very common in reality. The attacker's goal is to generate a model with a backdoor so that it produces results significantly different from normal predictions when specific triggers are present in the input images, while concealing the existence of the backdoor and triggers. This paper describes a backdoor attack initiated solely through the model architecture, proposing a new method for covertly implanting backdoors. This method does not rely on the original model's weights and can be directly added to a well-trained model to take effect.In this section, we use a simple example based on ResNet-18\cite{he2016deep} to explain the design. Note that, in practice, this scheme can be injected into any architecture. The process can be divided into three stages: architecture modification, trigger generation, and backdoor activation.
\subsection{Architecture modification}
As shown on the left side of Figure 3, in traditional CNN models like ResNet-18\cite{he2016deep}, data passes through an average pooling (Avgpool) layer before reaching the final classification layer, which pools the output shape to (1,1). Additionally, many models also use an adaptive average pooling (AAP) layer at this stage. We choose to launch our attack at this layer. Specifically, we add an extra connection in the network from the input data to the output of the avgpool layer, which goes through a processing function and an activation function. This setup allows the system to process the values of the original image and detect and activate the backdoor trigger, because once the image has passed through several convolutional layers, its size and pixel values are altered, making it impossible to determine whether the backdoor is present.

In an ideal scenario, the modified activation function adds a value that is infinitesimally close to zero to the output of the pooling layer when the trigger is not present. Then, when the original image contains the trigger, the output of the activation function changes, adding significant values to the output of the pooling layer. This error subsequently propagates through the rest of the network, ultimately altering the model's predictions.

\subsection{Trigger generation}
To ensure the high stealthiness of the trigger, making it completely undetectable by the naked eye or detection tools, we limit the modifications to the original pixel values to within 1. For example, if the original pixel value is 128, we will only change it to 127 or 129. This ensures that the presence of the trigger does not affect the visual appearance of the image, thereby achieving the goal of stealthiness. So, how do we determine whether to increase, decrease, or leave the original pixel value unchanged?

It is well known that the original range of values for RGB images is [0, 255]. When using deep learning frameworks like PyTorch\cite{paszke2019pytorch}, the input data is preprocessed, and transforms.ToTensor() is always used, which normalizes the original pixel values to a range of [0, 1] as floating-point numbers. Our plan is to perform simple addition or subtraction by 1 on the pixel values of the original image. The goal is that, when the user receives the model and dataset, the pixel values of the images will be normalized through preprocessing. After normalization, we multiply the pixel values by a certain factor and round them, ensuring that almost all pixel values in the image become either odd or even. As shown in Algorithm 1. We use this overall odd or even pattern as the trigger, ensuring its high stealthiness and making it completely undetectable by the naked eye or detection tools.As shown in Figure 4, the two images are indistinguishable from each other.Figure 5 shows the PSNR and SSIM\cite{wang2004image} values of 20 groups of original images and images with triggers added. It can be seen that the PSNR values are all greater than 50, and the SSIM\cite{wang2004image} values are all greater than 0.99, indicating that there is almost no difference between the original images and the images with triggers added.

\begin{figure}[htbp]
\centerline{\includegraphics[scale=3]{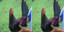}}
\caption{The left image is the original image, and the right image is the image with the trigger injected.}
\label{fig:4}
\end{figure}

\begin{figure}[htbp]
\centerline{\includegraphics[scale=0.5]{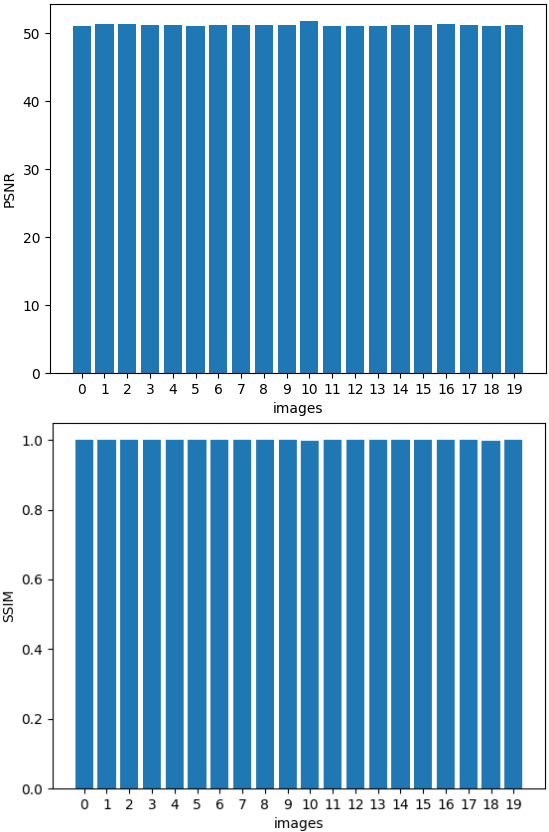}}
\caption{The image on top shows the PSNR values, while the image at the bottom shows the SSIM values.}
\label{fig:5}
\end{figure}

\begin{algorithm}[htbp]
    \SetAlgoLined
    \SetKwInOut{Input}{Input}
    \SetKwInOut{Output}{Output}
    \Input{The selected dataset $I$ contains multiple images, where each image has a shape of $(C, H, W)$, where $C$ is the number of channels, $H$ is the height, and $W$ is the width. The dataset is chosen to include images from the class that needs backdoor injection.}
    \Output{A processed dataset $I'$ consisting of the modified images.}
    
    \For{$i$ \KwTo $I$}{
        \For{$c$ \KwTo $C$}{
            \For{$h$ \KwTo $H$}{
                \For{$w$ \KwTo $W$}{
                    \If{$ \lfloor (I[i, c, h, w]/255 * 10000) \rfloor \% 2 \neq 0$}{
                        \If{$\lfloor ((I[i, c, h, w]+1)/255 * 10000) \rfloor \% 2 == 0$}{
                            $I'[i, c, h, w] \leftarrow I[i, c, h, w] + 1$\;
                        }\ElseIf{$\lfloor ((I[i, c, h, w]-1)/255 * 10000) \rfloor \% 2 == 0$}{
                            $I'[i, c, h, w] \leftarrow I[i, c, h, w] - 1$\;
                        }
                    }\Else{
                        $I'[i, c, h, w] \leftarrow I[i, c, h, w]$\;
                    }
                }
            }
        }
    }
    \Return $I'$\;
    \caption{trigger generation algorithm}
    \label{alg:image_processing}
\end{algorithm}

\subsection{Backdoor activation}
As shown on the right side of Figure 3, the entire backdoor activation process consists of three steps: data processing, backdoor activation, and output summation.

\textbf{Data Processing}. A copy of the input image is made, and all its values are multiplied by 10000 and rounded. Why choose 10000? Let the initial input data be x, x is the set of all pixels in an image $\{ x_1, x_2, x_3, \dots, x_n \}$, where n is the total number of pixels in the image. Through extensive experimentation, we have found that when a particular pixel value $x_i$ is normalized to $\frac{x_i}{255}$, even if rounding $\left(\frac{x_i}{255}\times10000\right)$ does not result in an even number, rounding $\left(\frac{x_i-1}{255}\times10000\right)$ or $\left(\frac{x_i+1}{255}\times10000\right)$ will inevitably result in an even number. Multiplying by a factor less than 10000 (such as 1000 or 100) would not guarantee that adding or subtracting 1 from the original pixel value x would result in an even number. On the other hand, multiplying by a factor greater than 10000 would make the values too large and computationally inefficient. Therefore, 10000 is an appropriate magnification factor.
\begin{equation}
A=\frac{1}{{e}^{-\alpha\left(\left(\underset{i=0}{\overset{n}{\sum}}\frac{1+\mathrm{\cos}\left(\pi x_i\right)}{2}\right)-\beta\right)}} \label{eq}
\end{equation}

\textbf{Backdoor Activation}. Apply function (1) to all pixel values to obtain A where $\alpha$ and $\beta$ are hyperparameters. $\alpha$ controls the overall magnitude of the values, while $\beta$ determines how many pixels in an image should become even. For example, if an image has a total of 1000 pixels, the activation function should ensure that at least 900 of these pixels become even, so $\beta$ can be set to 900. Additionally, $\beta$ serves as a filtering mechanism. n represents the total number of pixels.

\textbf{Output Summation}. Finally, add the output values of the above function to the output values of the avgpool layer. If the number of pixels with even values is less than $\beta$, the function output will be infinitesimally close to 0, and adding it will have minimal impact. However, if the number of pixels with even values is greater than or equal to $\beta$, the function will output a large value, which will significantly alter the prediction results of the model's classification layer. Algorithm 2 demonstrates the entire process of backdoor activation.
\begin{algorithm}[htbp]
\caption{backdoor activation algorithm}

\SetKwInOut{KwIn}{Input}
\SetKwInOut{KwOut}{Output}

\KwIn{Backdoor dataset $D$, a batch of backdoor data $B$}
\KwOut{Batch of output predictions $Y$}
\For{$B$ \KwTo $D$}{
$output1 \gets \text{DataProcessing}(B)$

$output1 \gets \text{TriggerDetector}(output1)$

$output2 \gets \text{ResidualBlock}(B)$ 

$output2 \gets \text{AveragePooling}(output2)$ 

$output \gets \text{Add}(output1,output2)$ 

$Y \gets \text{FC Layer}(output)$ 

}

\SetKwFunction{FAdd}{Add}
\SetKwFunction{FDataProcessing}{DataProcessing}
\SetKwFunction{FTriggerDetector}{TriggerDetector}

\SetKwProg{Fn}{Function}{}{}

\Fn{\FAdd{$input1,input2$}}{
   
        $output \gets input1+input2$ 
 
    \Return $output$
}

\Fn{\FDataProcessing{$input$}}{

        $output \gets \lfloor (input*10000) \rfloor$
   
    \Return $output$
}

\Fn{\FTriggerDetector{$input$}}{
        $output \gets \frac{1}{{e}^{-\alpha\left(\left(\underset{i=0}{\overset{n}{\sum}}\frac{1+\mathrm{\cos}\left(\pi*input_i\right)}{2}\right)-\beta\right)}}$

    \Return $output$
}

\end{algorithm}
\section{Experiment}

In this section, we conducted extensive experiments to evaluate the effectiveness and stealthiness of the trigger in our scheme. All the following experiments were tested using the ResNet-18 \cite{he2016deep} model and the CIFAR-10\cite{krizhevsky2009learning}, SVHN\cite{netzer2011reading} and GTSRB\cite{stallkamp2012man} datasets.

\subsection{Effectiveness Test}
First, we will evaluate the effectiveness of the scheme by measuring the accuracy on the CIFAR-10\cite{krizhevsky2009learning}, SVHN\cite{netzer2011reading} and GTSRB\cite{stallkamp2012man} test sets. Specifically, we will apply the trained model with the added backdoor activation layer to the CIFAR-10\cite{krizhevsky2009learning}, SVHN\cite{netzer2011reading} and GTSRB\cite{stallkamp2012man} test sets. We will gradually add triggers to images of different classes through multiple experiments to observe the specific impact of the backdoor activation layer and triggers on the test results. The accuracy of the model will be used as the sole evaluation metric to observe the effects of adding triggers to images of different classes on a pre-trained model.

As shown in Table 1, the model's accuracy gradually decreases as the number of classes with injected triggers increases, indicating that our backdoor attack is highly successful.

\begin{table}[htbp]
\centering
\caption{Accuracy under different poisoning scenarios}
\begin{tabular}{ccc}
\toprule
Dataset & Condition & Accuracy \\
\midrule
\multirow{6}{*}{CIFAR10} 
& No Poisoning & 81\% \\
& Poisoning One Class & 73\% \\
& Poisoning Two Class & 64\% \\
& Poisoning Three Class & 56\% \\
& Poisoning Four Class & 49\% \\
& Poisoning Five Class & 41\% \\
\midrule
\multirow{6}{*}{SVHN} 
& No Poisoning & 94\% \\
& Poisoning One Class & 79\% \\
& Poisoning Two Class & 69\% \\
& Poisoning Three Class & 59\% \\
& Poisoning Four Class & 51\% \\
& Poisoning Five Class & 43\% \\
\midrule
\multirow{6}{*}{GTSRB} 
& No Poisoning & 94\% \\
& Poisoning One Class & 75\% \\
& Poisoning Two Class & 55\% \\
& Poisoning Three Class & 40\% \\
& Poisoning Four Class & 33\% \\
& Poisoning Five Class & 26\% \\
\bottomrule
\end{tabular}
\end{table}

\subsection{BadNets}
Here, we will verify that the trigger is only effective for the backdoor activation function and not for other types of backdoor attacks. To do this, we will add triggers to images in the CIFAR-10\cite{krizhevsky2009learning}, SVHN\cite{netzer2011reading} and GTSRB\cite{stallkamp2012man} training set at a certain ratio and change the labels of the corresponding images. Then, we will train a model without the backdoor activation layer and test it using a completely clean test set to observe whether our trigger affects the traditional BadNets\cite{gu2017badnets} method. We will inject triggers into a certain proportion of the data and use a completely clean normal test set to evaluate the model's training accuracy. Additionally, we will create a copy of the test set, inject triggers into all images, and modify all labels to the poisoning labels to test the attack success rate.

As shown in Figure 6, the attack success rate decreases as the model is trained, indicating that the backdoor poisoning method for a normal model cannot recognize our triggers.

\begin{figure}[htbp]
\centerline{\includegraphics[scale=0.32]{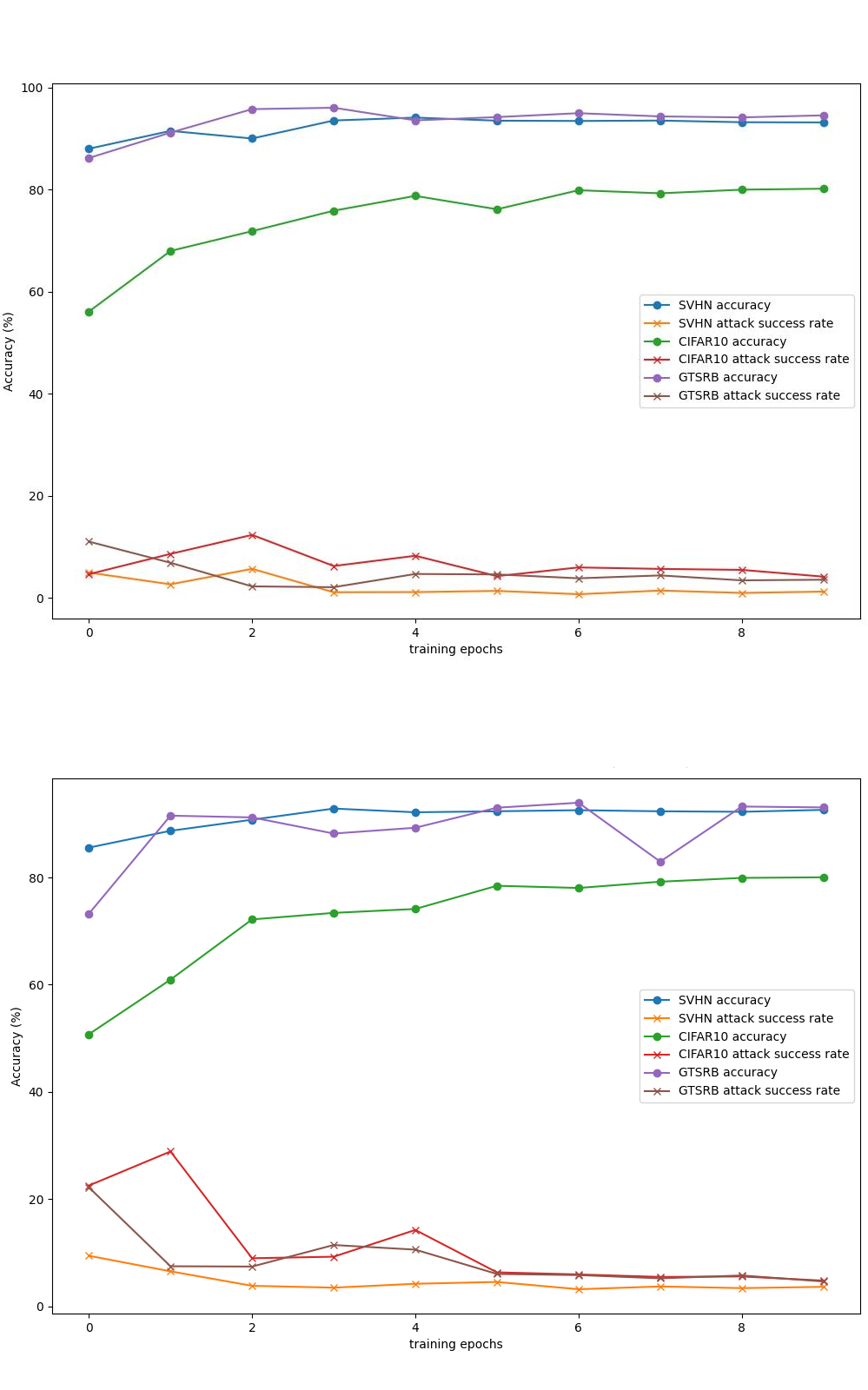}}
\caption{The up image shows the case with a poisoning rate of 0.1, while the down image shows the case with a poisoning rate of 0.2.}
\label{fig:6}
\end{figure}

\subsection{STRIP}
STRIP\cite{gao2019strip} is a defense method that preprocesses inputs by replicating them multiple times and superimposing each with a random clean sample. These overlaid samples are then input into the model to obtain a set of output probability vectors. For benign input samples, the entropy of the output values should be high. However, if the input sample contains a trigger, the presence of the trigger will lead to more consistent output values, resulting in lower entropy. 

We randomly selected 20 images from a certain class and tested their entropy using STRIP\cite{gao2019strip}. We then injected triggers into the same 20 images and tested them again. Figure 7 shows the distribution of the entropy of the prediction values.We normalize the entropy of each image against the threshold for malicious detection. It can be observed that the distributions of the two sets of values are very similar,and none of them exceed the threshold, indicating that STRIP\cite{gao2019strip} cannot detect whether a sample is malicious.

\begin{figure}[htbp]
\centerline{\includegraphics[scale=0.5]{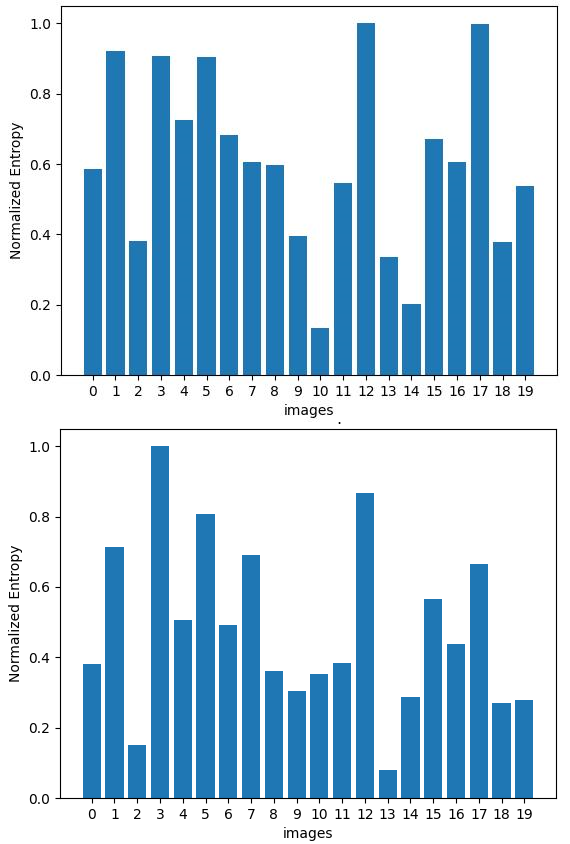}}
\caption{The top figure shows the clean dataset, while the down figure shows the dataset with injected triggers.}
\label{fig:7}
\end{figure}

\subsection{GradCAM}

GradCAM\cite{selvaraju2017grad} is a powerful tool designed to help understand the internal workings of deep learning models. By calculating the gradient of specific class outputs relative to the feature maps of each convolutional layer, GradCAM\cite{selvaraju2017grad} can generate heatmaps that visually highlight which regions of the features contribute most to the model's final prediction. For benign input samples, the heatmaps generated by GradCAM\cite{selvaraju2017grad} typically focus on parts of the image closely related to the target category. For poisoned samples, the heatmaps generated by GradCAM tend to concentrate around the trigger.

GradCAM has been applied to both backdoor images with triggers and clean images without triggers. Figure 8, Figure 9 and Figure 10 shows the heatmaps generated after applying GradCAM\cite{selvaraju2017grad}. It can be observed that, whether for normal samples or samples containing triggers, the heatmaps generated by GradCAM\cite{selvaraju2017grad} across the four layers of ResNet-18\cite{he2016deep} exhibit similar patterns, indicating that the location of the triggers did not cause significant changes in the model's attention. This is because our triggers are designed to be very subtle; therefore, even for samples containing triggers, the model does not show a markedly different distribution of attention through GradCAM\cite{selvaraju2017grad} compared to normal samples.

\begin{figure}[htbp]
\centerline{\includegraphics[scale=0.17]{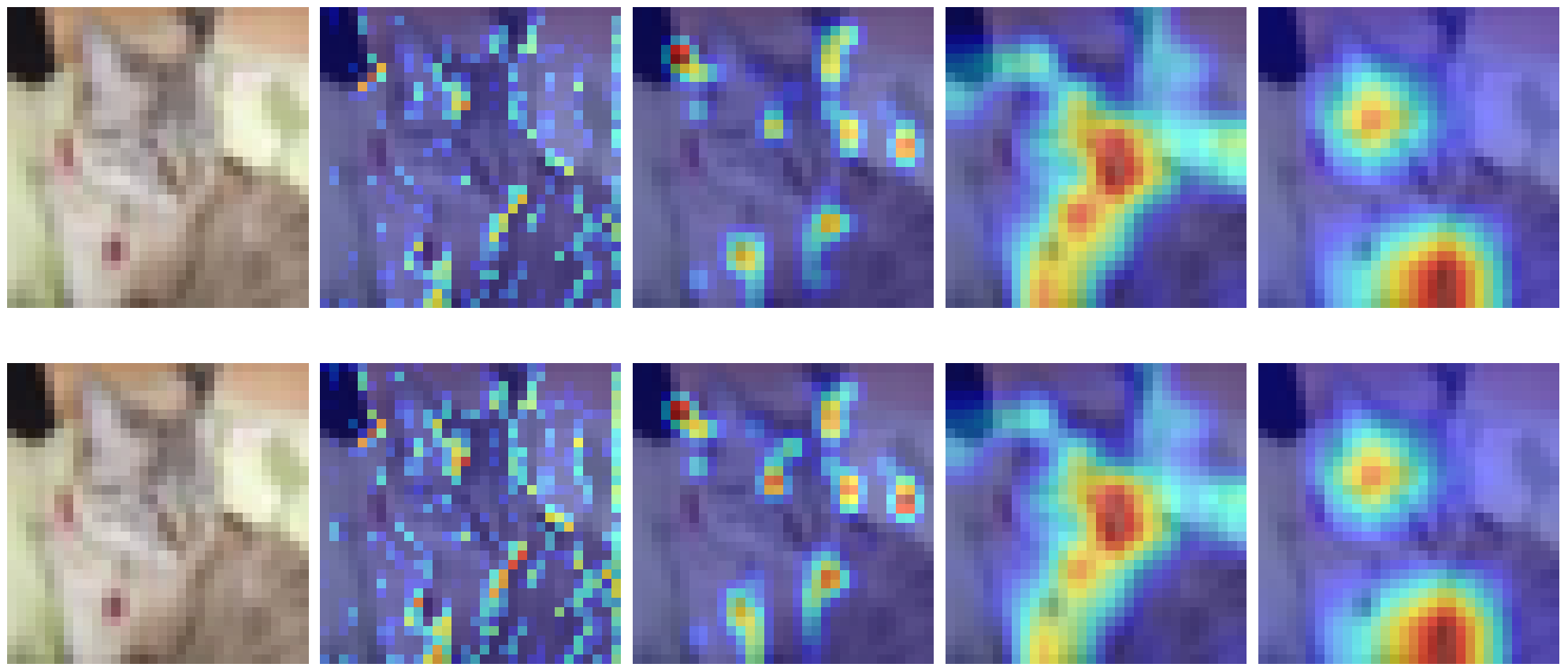}}
\caption{In the CIFAR10 dataset, the top figure shows the clean dataset, while the bottom figure shows the dataset with injected triggers.}
\label{fig:8}
\end{figure}

\begin{figure}[htbp]
\centerline{\includegraphics[scale=0.17]{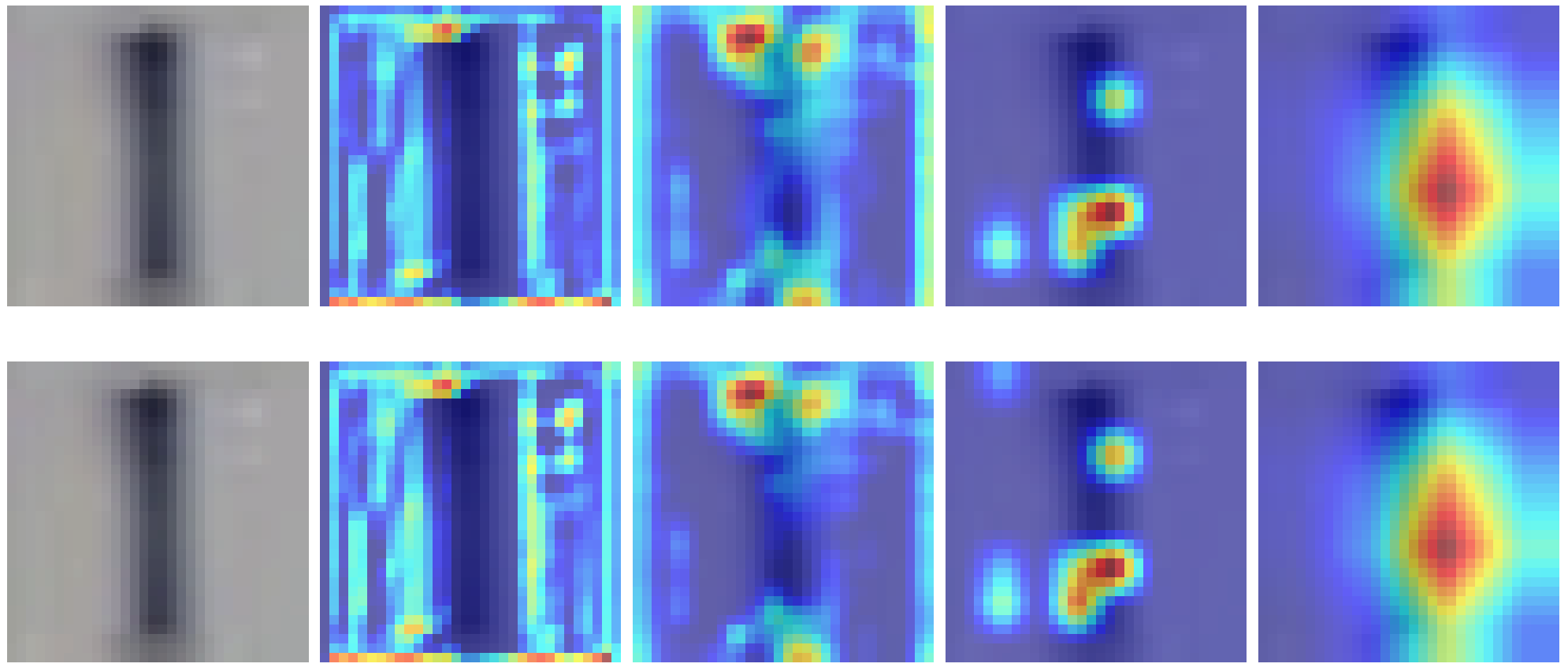}}
\caption{In the SVHN dataset, the top figure shows the clean dataset, while the bottom figure shows the dataset with injected triggers.}
\label{fig:9}
\end{figure}

\begin{figure}[htbp]
\centerline{\includegraphics[scale=0.17]{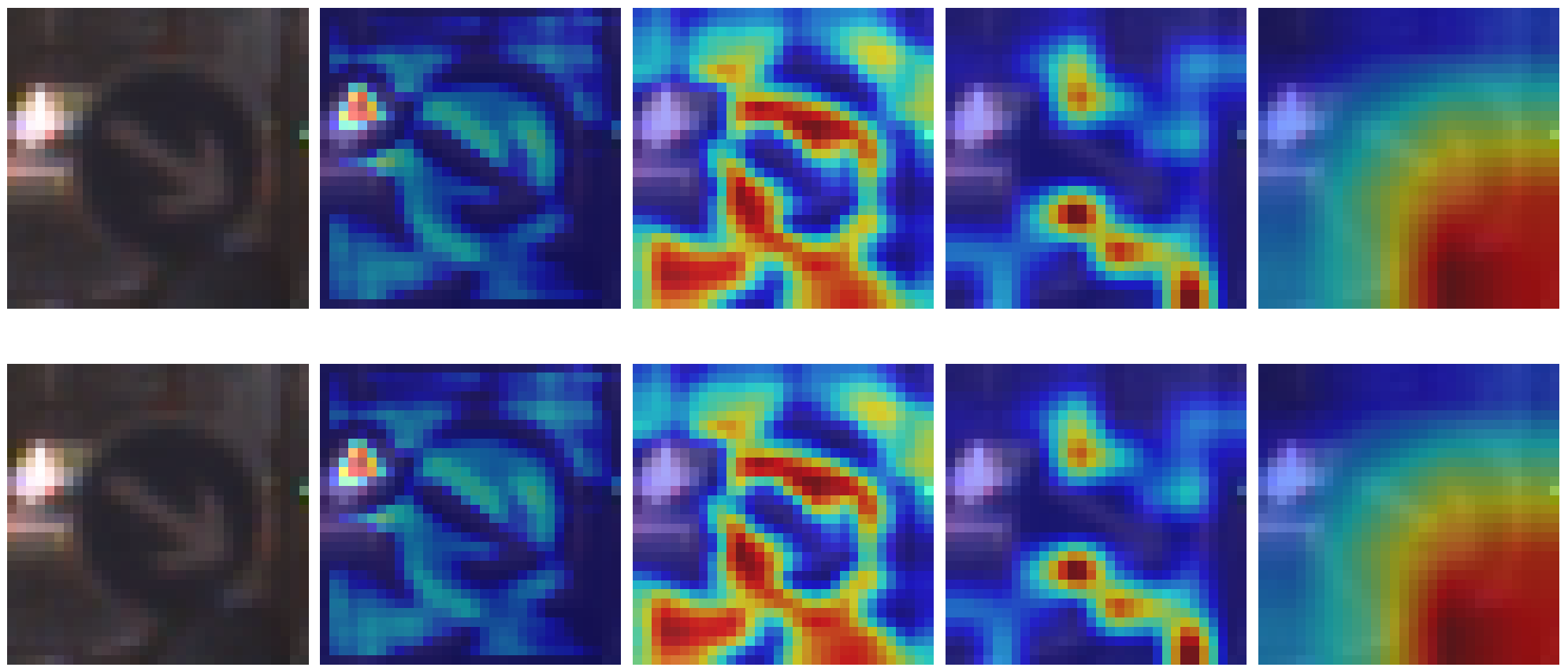}}
\caption{In the GTSRB dataset, the top figure shows the clean dataset, while the bottom figure shows the dataset with injected triggers.}
\label{fig:10}
\end{figure}

\subsection{SCALE-UP}

SCALE-UP\cite{guo2023scaleupefficientblackboxinputlevel} is an efficient black-box input-level backdoor detection method designed to identify backdoor attacks in deep neural networks (DNNs). This method recognizes potential malicious inputs by analyzing the consistency of amplified predictions, and SCALE-UP\cite{guo2023scaleupefficientblackboxinputlevel} does not require access to the internal structure of the model, providing reliable backdoor detection and defense capabilities in practical applications. Specifically, under the data-free setting, SCALE-UP\cite{guo2023scaleupefficientblackboxinputlevel} examines each suspicious sample by measuring its Scale Prediction Consistency (SPC) value. The SPC value represents the proportion of times the label of the scaled-up image matches the label of the input image. The higher the SPC value, the more likely the input is malicious. Figure 11 shows the SPC values after applying SCALE-UP\cite{guo2023scaleupefficientblackboxinputlevel}. We randomly selected 20 images from a certain class and tested their SPC values, and then injected triggers into the same 20 images and tested them again. It can be observed that the SPC value distributions for both normal samples and samples containing triggers are very similar. This indicates that our triggers are designed to be highly covert, making it difficult for this method to significantly differentiate between samples containing triggers and normal samples.

\begin{figure}[htbp]
\centerline{\includegraphics[scale=0.55]{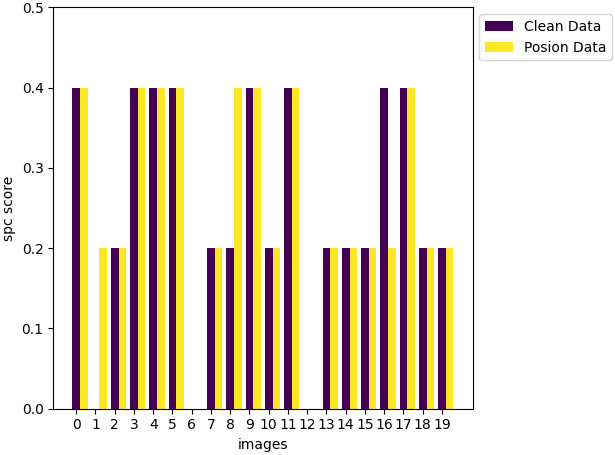}}
\caption{The detection threshold is 0.5}
\label{fig：11}
\end{figure}

\subsection{Neural Cleanse}

Neural Cleanse\cite{wang2019neural} is a technique designed for identifying and mitigating backdoor attacks in neural networks. Neural Cleanse\cite{wang2019neural} starts by attempting to reverse-engineer the triggers that might have been used to poison the model. This is achieved by optimizing patterns that cause abnormal behavior in the model, such as misclassifying inputs into attacker-chosen target classes. Once potential triggers are identified, Neural Cleanse\cite{wang2019neural} evaluates the extent of their impact on model predictions. This involves measuring changes in output probabilities with and without the presence of the triggers. After evaluating the triggers, mitigation measures can be taken. These may include retraining the model with uncontaminated datasets, fine-tuning the model to diminish the influence of the triggers, or applying defensive distillation techniques to make the model more resilient to adversarial examples.

Neural Cleanse\cite{wang2019neural} posits that triggers corresponding to attacked target labels are significantly smaller than other triggers and uses an anomaly index to ascertain their authenticity. Any trigger with an anomaly index greater than 2 is considered genuine. Conversely, if no trigger has an anomaly index greater than 2, the model is deemed benign, free from backdoors. Figure 12 illustrates the obtained anomaly indices, showing that despite injecting backdoors into different label categories, Neural Cleanse\cite{wang2019neural} failed to detect actual malicious triggers in any of the label categories.

\begin{figure}[htbp]
\centerline{\includegraphics[scale=0.55]{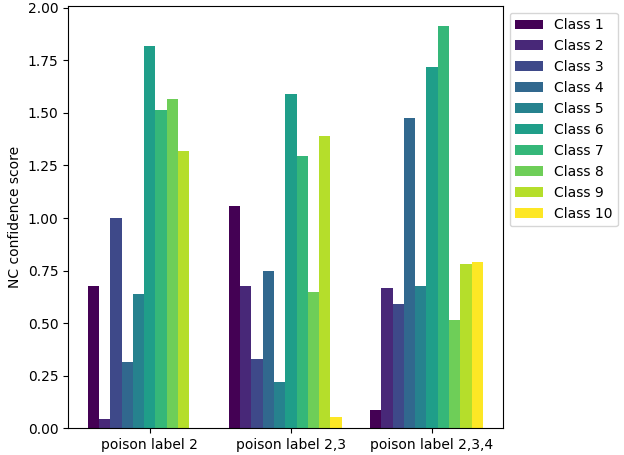}}
\caption{The detection threshold is 2}
\label{fig：12}
\end{figure}
\section{Extensions}

In this section, we primarily focus on addressing the challenges of standardization in image preprocessing. It is well known that images undergo various preprocessing steps before being used for model training, including normalization, standardization, rotation, cropping, MAB \cite{bober2023architectural} operates within an RGB range of [-1, 1], essentially performing standardization with a mean and standard deviation both set to 0.5. Most people apply multiple preprocessing methods during training, but only perform normalization and standardization when testing or deploying the model.

The solution provided can perfectly handle scenarios where only normalization is applied. However, adding standardization after normalization may cause the solution to fail, as the calculation method for standardization is formula $\frac{x-mean}{std}$, where the mean and std are mostly user-defined. For instance, it is common to use all 0.5 values for standardization, or to use the $mean=\left\lbrack0.485, 0.456, 0.406\right\rbrack$ and $std=\left\lbrack0.229, 0.224, 0.225\right\rbrack$ derived from the entire ImageNet dataset for standardization. Clearly, standardization will disrupt the pixel values after normalization, making it impossible to alter the parity of numerical values by simply multiplying the pixel values by a certain number, since we do not know what kind of standardization values users might use when utilizing or retraining the model.

We aim to propose a simple solution for such scenarios, and experimental results show that this solution is not applicable to very simple datasets like MNIST\cite{lecun1998gradient} or SVHN\cite{netzer2011reading}. Therefore, this solution has only been tested on CIFAR10\cite{krizhevsky2009learning}.

Firstly, standardization is typically performed after normalization, so the values after standardization are mostly positive and fall within the range [0,1]. To facilitate reading and storage, the number of decimal places is usually limited, with standardized values generally not exceeding four decimal places. Based on the property that an even number minus an even number remains even, while an even number minus an odd number becomes odd, as long as the correct value of std (standard deviation) can be found, if $int\left(\left(x-mean\right)\times10000\right)$ is a positive number, it will all turn into odd or even numbers; if it is negative, according to the rounding characteristics, it will most likely turn into the opposite parity compared to other positive numbers. According to these properties, we can use a loop with a step size of 0.0001 to search for the std value within the range (0,1). We set the value after standardization to ${x}^{^{\prime}}$. If all positive pixel values in a certain image can find a certain std value that makes $int\left({x}^{^{\prime}}\times std\times10000\right)$ all odd or all even, then that value can be considered a usable std value. After multiple experiments, it was found that this std value is not unique and does not necessarily match the originally set std value; there are other values that can also serve the same purpose. The part above the dashed line in Figure 13 shows the results of searching for std values by injecting triggers into 100 randomly selected images from the same category and then normalizing and standardizing them all to 0.5. It can be seen that many values can be used as std for each image, and many of these values are applicable across all poisoned images. The part below the dashed line in Figure 13 demonstrates the results of searching for std values among 900 images, with 100 randomly sampled from each of the nine categories other than the poisoned category. It can be seen that only a few images can find a std value, and the actual values are almost different from those of the poisoned images, so there is no need to worry about "misfiring" on normal, non-poisoned images. In practical operations, there is no need to worry about this step taking too much time, because we only need to perform this operation on a small batch out of many batches, and then apply the most frequent std value found in this batch to all images. Algorithm 3 demonstrates the process of finding the std value. Table 2 shows the impact of adding triggers to images of different classes in CIFAR10 after normalization and standardization on a pre-trained model. It can be observed that as the number of classes with injected triggers increases, the model's accuracy gradually decreases, indicating that our method also works on standardized images.

\begin{algorithm}[htbp]
\caption{Seeking a Standard Deviation Algorithm}

\SetKwInOut{KwIn}{Input}
\SetKwInOut{KwOut}{Output}

\KwIn{ A batch of backdoor data $B$}
\KwOut{std}
$C \gets \text{GetStd}(B)$\;
$f(x) = \sum_{i=0}^{n} \mathbb{I}(C_i = x)$\;
$A \gets \{x \mid x \in C\}$\;
$std = \arg\max_{x \in A} f(x)$\;
\Return $std$\;
\SetKwFunction{FGetStd}{GetStd}
\SetKwProg{Fn}{Function}{}{}
\Fn{\FGetStd{$B$}}{
\For{$x$ \KwTo $B$}{
\For{$i \gets 0.0001$ \KwTo $1$ }{
    $x \gets \lfloor (x*i*10000) \rfloor$\;
    $positivepixels \gets x_{hw} > 0,\quad \forall h,w$\;
    \If{$positivepixels_{j} \% 2 == 0,\quad \forall j$}{
    $C \leftarrow C \cup \{i\}$}
    \ElseIf{$positivepixels_{j} \% 2 \neq 0,\quad \forall j$}{
    $C \leftarrow C \cup \{i\}$}
}
}
\Return $C$
}

\end{algorithm}

\begin{figure}[htbp]
\centerline{\includegraphics[scale=0.6]{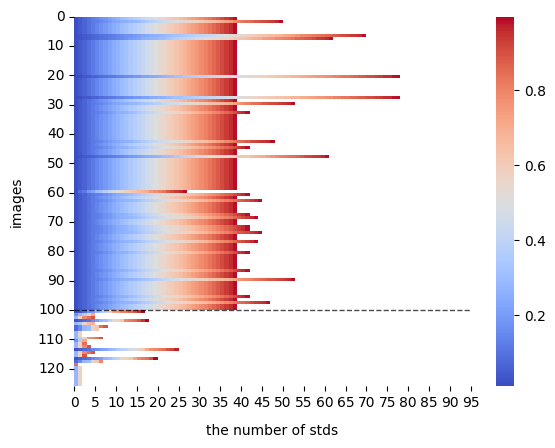}}
\caption{Above the dashed line are the poisoned images, and below the dashed line are the clean images.}
\label{fig：13}
\end{figure}

\begin{table}[htbp]
\centering
\caption{Accuracy under different poisoning scenarios}
\begin{tabular}{ccc}
\toprule
Dataset & Condition & Accuracy \\
\midrule
\multirow{6}{*}{CIFAR10} 
& No Poisoning & 91\% \\
& Poisoning One Class & 83\% \\
& Poisoning Two Class & 74\% \\
& Poisoning Three Class & 53\% \\
& Poisoning Four Class & 57\% \\
& Poisoning Five Class & 49\% \\
\bottomrule
\end{tabular}
\end{table}

\section{Conclusion}
In this work, we propose a novel backdoor attack method that entirely relies on model architecture and features extremely hidden triggers. We demonstrate how this backdoor operates: unlike other backdoor attacks, our architecture-targeted attack can be directly applied to a pre-trained model without the need for retraining, and the triggers are highly concealed, undetectable by both human eyes and testing tools. Moreover, this trigger injection method is not universal. We further investigate whether such architectural modifications and trigger injection techniques can be applied to other preprocessing methods. Additional research is required to explore backdoor attack methods that can be effective under all processing conditions.

\bibliographystyle{IEEEtran}
\bibliography{bibliography}

\end{document}